\documentclass{PoS}
\usepackage{amsmath,amssymb,epsfig,fleqn,url,psboxit,color,here,graphics,graphicx,rotating,multirow,wrapfig}
\usepackage{bigstrut}
\usepackage{booktabs}

\newcommand{\slim}{\mskip 1.5mu}              

\newcommand{\phiH}{\phi _h}
\newcommand{\phiS}{\phi _S}

\newcommand{\red}{\textcolor[rgb]{1.00,0.00,0.00}}
\newcommand{\blue}{\textcolor[rgb]{0.00,0.00,1.00}}
\newcommand{\green}{\textcolor[rgb]{0.00,0.59,0.00}}

\title{SIDIS transverse spin azimuthal asymmetries at COMPASS: Multidimensional analysis}

\ShortTitle{SIDIS TSAs at COMPASS: Multi\--D analysis}

\author{\speaker{Bakur Parsamyan}\thanks{on behalf of the COMPASS collaboration.}\\
        University of Turin and Torino Section of INFN\\
 Via P. Giuria 1, 10125 Torino, Italy\\
        E-mail: \email{bakur.parsamyan@cern.ch}}

\abstract{Exploration of transverse spin structure of the nucleon via study of the spin (in)dependent azimuthal asymmetries
in semi-inclusive deep inelastic scattering (SIDIS) and Drell-Yan (DY) reactions is
one of the main aspects of the broad physics program of the COMPASS experiment (CERN, Switzerland).
In past decade COMPASS has collected a considerable amount of polarized deuteron and proton SIDIS data, while recent 2014 and 2015 runs
were dedicated to the Drell-Yan measurements.
Results on SIDIS azimuthal effects provided so far by COMPASS play an important role in general understanding
of the three-dimensional nature of the nucleon. Giving access to the entire "twist-2" set of transverse momentum dependent
(TMD) parton distribution functions (PDFs) and fragmentation functions (FFs) COMPASS data are being widely used in phenomenological analyses
and experimental data fits. Recent unique and first ever x-$Q^{2}$-z-pT
multidimensional results for transverse spin asymmetries obtained by COMPASS serve as a direct and unprecedented
input for one of the hottest topics in the field of spin-physics: the TMD $Q^{2}$-evolution related studies.	
In addition, extraction of the Sivers and all other azimuthal effects from first ever polarized Drell-Yan data collected recently by COMPASS will reveal another
side of the spin-puzzle clarifying the link between SIDIS and Drell-Yan branches. This will be a unique possibility to test
predicted universality and key-features of TMD PDFs using essentially the same experimental setup and exploring the same kinematical domain. 	
In this review main focus will be given to the very recent
results from COMPASS multi-dimensional analysis of transverse spin asymmetries and to the physics aspects of
COMPASS polarized Drell-Yan program.}

\FullConference{QCD Evolution 2015 -QCDEV2015-\\
		26-30 May 2015\\
		Jefferson Lab (JLAB), Newport News Virginia, USA}

\begin{document}
%
%
%
%
%
%
%
%
\section{Introduction}	
\begin{wrapfigure}{r}{0.4\textwidth}
  \begin{center}
    \includegraphics[width=0.4\textwidth]{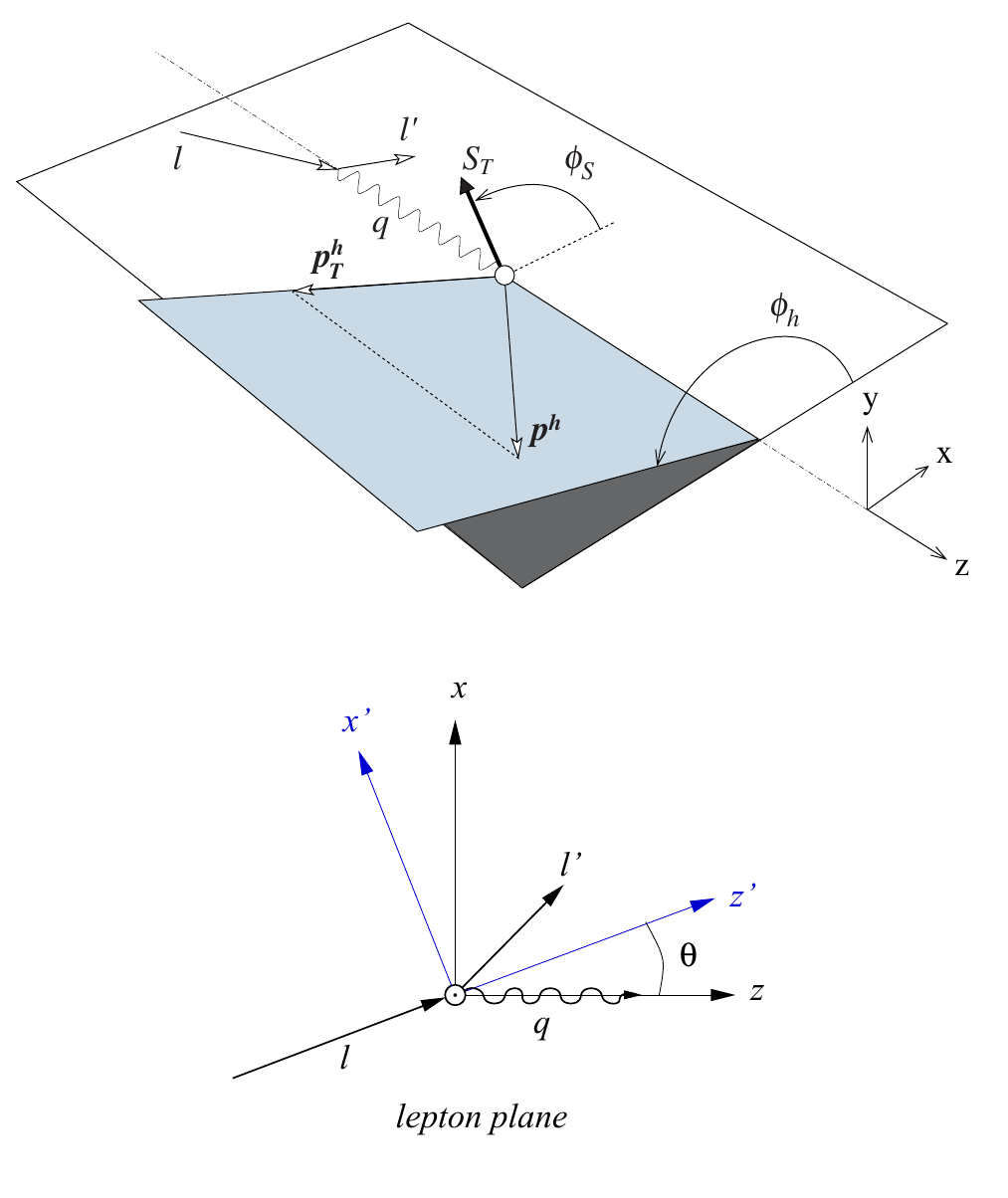}
  \end{center}
\caption{SIDIS process framework. 
}
  \label{fig:SIDISangles}
\end{wrapfigure}

Schematic view of the SIDIS framework and some notations and
definitions adopted in this letter such as coordinate system, azimuthal angles, etc. are presented in Figure~\ref{fig:SIDISangles}.
In this framework the target transverse polarization ($S_T$) is defined relative to the virtual photon momentum direction, which is the most
natural basis from the theoretical point of view. However, in experiment target is being polarized in laboratory system and
transverse polarization ($P_T$) is defined relative to the beam (incoming lepton) direction. As it was demonstrated in \cite{Kotzinian:1994dv}-\cite{Diehl:2005pc} this difference
influences azimuthal distributions in the final state. After applying appropriate conversions, the model-independent expression for the
SIDIS cross-section for transversely (w.r.t. lepton beam) polarized target can be re-written in the following way \cite{Kotzinian:1994dv}-\cite{Diehl:2005pc}:
{\small
\begin{eqnarray}
&& \hspace*{-1.0cm}\frac{{d\sigma }}{{dxdydzp_{T}^{h}dp_{T}^{h}d{\phiH}d{\phiS}}} = 2\green{\underline {\left[ {\frac{{\cos \theta }}{{1 - {{\sin }^2}\theta {{\sin }^2}{\phiS}}}} \right]}} \left[ {\frac{\alpha }{{xy{Q^2}}}\frac{{{y^2}}}{{2\left( {1 - \varepsilon } \right)}}\left( {1 + \frac{{{\gamma ^2}}}{{2x}}} \right)} \right]\left( {{F_{UU,T}} + \varepsilon {F_{UU,L}}} \right) \\ \nonumber
&&\hspace*{-0.9cm} \times\Bigg\{ 1 + \sqrt {2\varepsilon \left( {1 + \varepsilon } \right)} \blue{A_{UU}^{\cos {\phiH}}}\cos {\phiH} + \varepsilon \red{A_{UU}^{\cos \left( {2{\phiH}} \right)}}\cos \left( {2{\phiH}} \right) + \lambda \sqrt {2\varepsilon \left( {1 - \varepsilon } \right)} \blue{A_{LU}^{\sin {\phiH}}}\sin {\phiH}\\ \nonumber
&&\hspace*{-0.6cm}+\,\frac{{{P_{\rm{T}}}}}{{\green{\underline {\sqrt {1 - {{\sin }^2}\theta {{\sin }^2}{\phiS}} }} }}\Bigg[\left( {\green{\underline {\cos \theta }} \red{A_{UT}^{\sin \left( {{\phiH} - {\phiS}} \right)}} + \green{\underline {\frac{1}{2}\sin \theta \sqrt {2\varepsilon \left( {1 + \varepsilon } \right)} A_{UL}^{\sin {\phiH}}} }} \right)\sin \left( {{\phiH} - {\phiS}} \right)\\ \nonumber
&&\hspace*{3.0cm}+\,\left( {\green{\underline {\cos \theta }} \varepsilon \red{A_{UT}^{\sin \left( {{\phiH} + {\phiS}} \right)}} + \green{\underline {\frac{1}{2}\sin \theta \sqrt {2\varepsilon \left( {1 + \varepsilon } \right)} A_{UL}^{\sin {\phiH}}} }} \right)\sin \left( {{\phiH} + {\phiS}} \right)\\ \nonumber
&&\hspace*{3.0cm}+\,\,\green{\underline {\cos \theta }} \varepsilon \red{A_{UT}^{\sin \left( {3{\phiH} - {\phiS}} \right)}}\sin \left( {3{\phiH} - {\phiS}} \right)\\ \nonumber
&&\hspace*{3.0cm}+\,\,\green{\underline {\cos \theta }} \sqrt {2\varepsilon \left( {1 + \varepsilon } \right)} \blue{A_{UT}^{\sin {\phiS}}}\sin {\phiS}\\ \nonumber
&&\hspace*{3.0cm}+\,\left( {\green{\underline {\cos \theta }} \sqrt {2\varepsilon \left( {1 + \varepsilon } \right)} \blue{A_{UT}^{\sin \left( {2{\phiH} - {\phiS}} \right)}} + \green{\underline {\frac{1}{2}\sin \theta \varepsilon A_{UL}^{\sin 2{\phiH}}} }} \right)\sin \left( {2{\phiH} - {\phiS}} \right)\\ \nonumber
&&\hspace*{3.0cm}+\,\green{\underline {\frac{1}{2}\sin \theta \varepsilon A_{UL}^{\sin 2{\phiH}}\sin \left( {2{\phiH} + {\phiS}} \right)}} \Bigg]\\ \nonumber
&&\hspace*{-0.6cm}+\,\frac{{{P_{\rm{T}}}\lambda }}{\green{\underline {\sqrt {1 - {{\sin }^2}\theta {{\sin }^2}{\phiS}} } }}\Bigg[\left( {\green{\underline {\cos \theta }} \sqrt {\left( {1 - {\varepsilon ^2}} \right)} \red{A_{LT}^{\cos \left( {{\phiH} - {\phiS}} \right)}} + \green{\underline {\frac{1}{2}\sin \theta \sqrt {2\varepsilon \left( {1 - \varepsilon } \right)} A_{LL}^{\cos {\phiH}}} }} \right)\cos \left( {{\phiH} - {\phiS}} \right)\\ \nonumber
&&\hspace*{3.0cm}+\,\left( {\green{\underline {\cos \theta }} \sqrt {2\varepsilon \left( {1 - \varepsilon } \right)} \blue{A_{LT}^{\cos {\phiS}}} + \green{\underline {\sin \theta \sqrt {\left( {1 - {\varepsilon ^2}} \right)} {A_{LL}}} }} \right)\cos {\phiS}\\ \nonumber
&&\hspace*{3.0cm}+\,\left( {\green{\underline {\cos \theta }} \sqrt {2\varepsilon \left( {1 - \varepsilon } \right)} \blue{A_{LT}^{\cos \left( {2{\phiH} - {\phiS}} \right)}}} \right)\cos \left( {2{\phiH} - {\phiS}} \right)\\ \nonumber
&&\hspace*{3.0cm}+\,\green{\underline {\frac{1}{2}\sin \theta \sqrt {2\varepsilon \left( {1 - \varepsilon } \right)} A_{LL}^{\cos {\phiH}}\cos \left( {{\phiH} + {\phiS}} \right)}} \Bigg]\Bigg\}
\label{eq:SIDIS}
\end{eqnarray}
}
where $\varepsilon = (1-y -\frac{1}{4}\slim \gamma^2 y^2)/(1-y +\frac{1}{2}\slim y^2 +\frac{1}{4}\slim \gamma^2
y^2)$ and $\gamma = 2 M x/Q$ and $\theta$ is the angle between beam and virtual photon momentum directions (see Figure \ref{fig:SIDISangles}).
When compared with the "standard" cross-section \cite{Parsamyan:2015myq},
in which the effects due to the $P_T$ to $S_T$ transition have been neglected, \ref{eq:SIDIS} contains new
\footnote{all "new" terms are underlined and marked in green.}
$sin\theta$-scaled terms and $\theta$-depending factors and two extra azimuthal modulations
($\sin \left( {2{\varphi _h} + {\varphi _S}} \right)$ and $\cos \left( {{\varphi _h} + {\varphi _S}} \right)$) related to longitudinal amplitudes
(these two amplitudes have been measured to be compatible with zero and are not going to be discussed in this letter).

Beside those two terms, the expression \ref{eq:SIDIS} counts in total eight more $w_i(\phiH, \phiS)$ azimuthal modulations.
Each of this effects leads to a $A_{BT}^{w_i(\phiH, \phiS)}$ Transverse-Spin-dependent Asymmetry (TSA) defined as a ratio of
the associated structure function $F_{BT}^{w_i(\phiH,\phiS)}$ to the azimuth-independent one
${F_{UU}}={{F_{UU,T}} + \varepsilon {F_{UU,L}}}$. Here the superscript of the asymmetry
indicates corresponding modulation, the first and the second subscripts - respective ("U"-unpolarized, "L"-longitudinal and
"T"-transverse) polarization of beam and target. Five amplitudes which depend only on target polarization are the target Single-Spin
Asymmetries (SSA), the other three
which depend also on $\lambda$ beam longitudinal polarization are known as Double-Spin Asymmetries (DSA).

As it can be seen from \ref{eq:SIDIS} in several cases TSAs are being mixed with $sin\theta$-scaled longitudinal amplitudes.
Since $sin\theta$ is a rather small quantity in COMPASS kinematics \cite{Parsamyan:2013ug} the influence of the additional to the TSAs terms,
represented by $sin\theta$-scaled longitudinal-spin amplitudes and $\theta$-angle dependent factors,
is sizable only in the case of $A_{LT}^{\cos {\varphi _S}}$ DSA, which, even taking into account
the smallness of $sin\theta$, is still sizably affected by a large $A_{LL}$ amplitude \cite{Alekseev:2010ub}.
To correct the $A_{LT}^{\cos {\varphi _S}}$ asymmetry we have used the $A_{LL}$ values evaluated based on \cite{Anselmino:2006yc})
which are in close agreement with the data \cite{Parsamyan:2013ug,Alekseev:2010ub}.
%
%
%
%

In the QCD parton model approach four of the eight transverse spin asymmetries ($A_{UT}^{sin(\phi_h-\phi_S)}$,
$A_{UT}^{sin(\phi_h+\phi_S)}$, $A_{UT}^{\sin (3\phi _h -\phi _s )}$ SSAs and $A_{LT}^{\cos (\phi _h -\phi _s )}$ DSA)
have Leading Order (LO) or leading-twist interpretation. The first two are the "Sivers" and "Collins" effects
\cite{Adolph:2014zba}--\cite{Adolph:2012sp} which are the most studied ones. These asymmetries are given as convolutions of:
 $f_{1T}^{\perp q}$ Sivers PDF with $D_{1q}^h$ ordinary FF, and $h_{1}^{q}$ "transversity" PDF with the $H_{1q}^{\perp h}$ Collins FF, respectively.
 The other two LO terms are the $A_{UT}^{\sin(3\phiH -\phiS )}$
 single-spin asymmetry (related to $h_{1T}^{\perp\,q}$ ("pretzelosity") PDF
 \cite{Parsamyan:2014uda}--\cite{Parsamyan:2007ju}) and $A_{LT}^{\cos (\phiH -\phiS )}$ DSA (related to $g_{1T}^q$
("worm-gear") distribution function \cite{Parsamyan:2014uda}--\cite{Kotzinian:2006dw},\cite{Anselmino:2006yc}).

Remaining four asymmetries 
are so-called "higher-twist" effects\footnote{in equations \ref{eq:SIDIS}--\ref{eq:DY} the twist-2 amplitudes are marked in red and
higher-twist ones in blue}.
Corresponding structure functions enter at sub-leading order ($Q^{-1}$) and contain
terms given as various mixtures of twist-two and twist-three (induced by quark-gluon correlations) parton distribution
and fragmentation functions \cite{Mulders:1995dh}--\cite{Mao:2014fma}. However, applying wildly used
"Wandzura-Wilczek approximation" this higher twist objects can be simplified to the twist-two order (see \cite{Mulders:1995dh,Bacchetta:2006tn} for more details).
Complete list of "twist-two"-level interpretations for all eight TSAs is quoted in \ref{eq:asy_interp}.
{\small
\begin{align}\label{eq:asy_interp}
\red{A_{UT}^{\sin (\phiH -\phiS )}} \propto f_{1T}^{\bot q} \otimes
D_{1q}^h,\ \ &
\red{A_{UT}^{\sin (\phiH +\phiS )}} \propto h_1^q \otimes H_{1q}^{\bot
h},  \\ \nonumber
\red{A_{UT}^{\sin (3\phiH
-\phiS )}} \propto h_{1T}^{\bot q} \otimes H_{1q}^{\bot
h},\ \ & \red{A_{LT}^{\cos (\phiH -\phiS )}} \propto g_{1T}^q \otimes
D_{1q}^h \\ \nonumber
\blue{A_{UT}^{\sin (\phiS )}} \propto {Q}^{-1}({h_1^q \otimes
H_{1q}^{\bot h} +f_{1T}^{\bot q} \otimes D_{1q}^h }),\ &
\blue{A_{UT}^{\sin (2\phiH -\phiS )}} \propto
{Q}^{-1}({h_{1T}^{\bot q} \otimes H_{1q}^{\bot h}
+f_{1T}^{\bot q} \otimes D_{1q}^h }),\ \\ \nonumber
\blue{A_{LT}^{\cos (\phiS )}} \propto {Q}^{-1}(g_{1T}^q \otimes
D_{1q}^h),\ \ &
\blue{A_{LT}^{\cos (2\phiH -\phiS )}} \propto {Q}^{-1}
(g_{1T}^q \otimes D_{1q}^h).
\end{align}
}
The whole set of eight SIDIS asymmetries has been already measured at COMPASS for both deuteron and proton targets
(See \cite{Adolph:2014zba}--\cite{Parsamyan:2007ju} and references therein).
%
%
%
%
%
%
%
%
%

\begin{wrapfigure}{r}{0.4\textwidth}
  \begin{center}
    \includegraphics[width=0.4\textwidth]{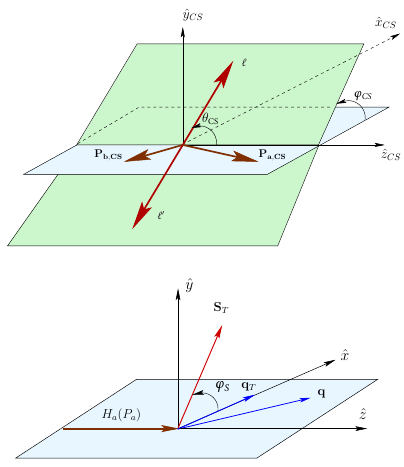}
  \end{center}
\caption{Drell-Yan process framework. 
}
  \label{fig:DYangles}
\end{wrapfigure}

Using similar notations, single-polarized ($\pi N^\uparrow$) Drell-Yan cross-section at leading order can be written in the following
%
%
model-independent way \cite{Gautheron:2010wva}:
{\small
\begin{eqnarray}
  &&\hspace*{-1.0cm}\frac{{d{\sigma ^{LO}}}}{{d\Omega }} = \frac{{\alpha _{em}^2}}{{F{q^2}}}F_U^1 \left\{ {1 + {{\cos }^2}\theta  + {{\sin }^2}\theta \red{A_U^{\cos 2{\varphi _{CS}}}}\cos 2{\varphi _{CS}}} \right. +
  {S_T}\left[ {\left( {1 + {{\cos }^2}\theta } \right)\red{A_T^{\sin {\varphi _S}}}\sin {\varphi _S}} \right.\hfill \\ \nonumber
  &&\hspace*{+1.9cm}{\text{   }}  + {\sin ^2}\theta \left(\red{A_T^{\sin \left( {2{\varphi _{CS}} + {\varphi _S}} \right)}}\sin \left( {2{\varphi _{CS}} + {\varphi _S}} \right)
   + \left. {\left. { \red{A_T^{\sin \left( {2{\varphi _{CS}} - {\varphi _S}} \right)}}\sin \left( {2{\varphi _{CS}} - {\varphi _S}} \right)} \right]}\right) \right\}.
\label{eq:DY}
\end{eqnarray}
}
Here $\varphi_{CS}$ and $\varphi_S$ angular variables are defined in "Collins-Soper" and "target rest" frames, correspondingly (see Figure~\ref{fig:DYangles}).
Similarly to the SIDIS case, "U", "L" and "T" subscripts denote the state of the target polarization while the superscript
indicates the corresponding modulation.

As one can see, at LO the DY cross-section contains only one unpolarized and three target transverse spin dependent azimuthal asymmetries.
Within the same QCD parton model approach, at variance with the SIDIS-case,
Drell-Yan asymmetries are interpreted as convolutions of two TMD PDFs, one of the the "beam" and one of the "target" hadron.
Quoting only the target nucleon PDFs: the $A_{T}^{\sin \varphi _s }$, $A_{T}^{\sin
(2\varphi _{CS} -\varphi _s )}$ and $A_{T}^{\sin(2\varphi _{CS} +\varphi _s )}$ give access to the
"Sivers" $f_{1T}^{\perp\,q}$, "transversity" $h_1^q$ and "pretzelosity" $h_{1T}^{\perp\,q}$, distribution functions, respectively.
Within the QCD-concept of \textit{generalized universality} of TMD PDFs it appears that same parton distribution functions can be accessed
both in SIDIS and Drell-Yan (see the Table.~\ref{tab:PDFs} for the complete list).
Therefore, future COMPASS results on DY asymmetries are intriguingly complementary to the results previously obtained
by the same collaboration for azimuthal effects in SIDIS.
The comparison of two sets will give an unprecedented opportunity to access TMD PDFs via two mechanisms and test their universality and key features
(for instance, predicted Sivers and Boer-Mulders PDFs sign change) using essentially same experimental setup.

{\small
\begin{table}[H]
\centering
%
\begin{tabular}{@{}ccc@{}} 
  \bottomrule
  SIDIS $\ell^\rightarrow N^\uparrow$ & TMD PDF & DY $\pi N^\uparrow$ (LO)\bigstrut\\
  \midrule
  \red{$A_{UU}^{\cos 2\phi _h}$}, \blue{$A_{UU}^{\cos \phi _h}$} & $h_{1}^{\bot q}$& \red{$A_{U}^{\cos 2\varphi _{CS}}$} \bigstrut\\
  \midrule
  \red{$A_{UT}^{\sin (\phi _h -\phi _s )}$}, \blue{$A_{UT}^{\sin\phi _s}$}, \blue{$A_{UT}^{\sin (2\phi _h -\phi _s )}$} & $f_{1T}^{\bot q}$& \red{$A_{T}^{\sin \varphi _{S}}$} \bigstrut\\
  \midrule
  \red{$A_{UT}^{\sin (\phi _h +\phi _s -\pi)}$}, \blue{$A_{UT}^{\sin\phi _s}$} & $h_{1}^{q}$& \red{$A_{T}^{\sin (2\varphi _{CS} -\varphi _{S} )}$} \bigstrut\\
  \midrule
  \red{$A_{UT}^{\sin (3\phi _h -\phi _s )}$}, \blue{$A_{UT}^{\sin (2\phi _h -\phi _s )}$} & $h_{1T}^{\bot q}$& \red{$A_{T}^{\sin (2\varphi _{CS} +\varphi _{S} )}$} \bigstrut\\
  \midrule
  \red{$A_{LT}^{\cos (\phi _h -\phi _s )}$}, \blue{$A_{LT}^{\cos\phi _s}$}, \blue{$A_{LT}^{\cos (2\phi _h -\phi _s )}$} & $g_{1T}^{ q}$& \red{double-polarized DY} \bigstrut\\
 \bottomrule
\end{tabular}
\caption{\label{tab:PDFs}Nucleon TMD PDFs accessed via SIDIS and Drell-Yan azymmetries.}
\end {table}
}
Another hot topic being addressed by the COMPASS collaboration is the multi-differential analysis of SIDIS data.
In general, asymmetries being represented as convolutions of different TMD distribution functions are considered to be complex objects \textit{a priori}
dependent on the experimental choice of multidimensional kinematical phase-space. Thus, in order to reveal the most complete
multivariate dependence of TMD PDFs, it is important to extract azimuthal amplitudes as multi-differential functions of kinematical variables.
In practice, available experimental data often are too limited for such an ambitious approach
and studying dependence of the asymmetries on some specific kinematic variable one is forced to stick to one-dimensional case integrating over all the
other variables.
\begin{figure}[H]
\center
\includegraphics[width=0.9\textwidth]{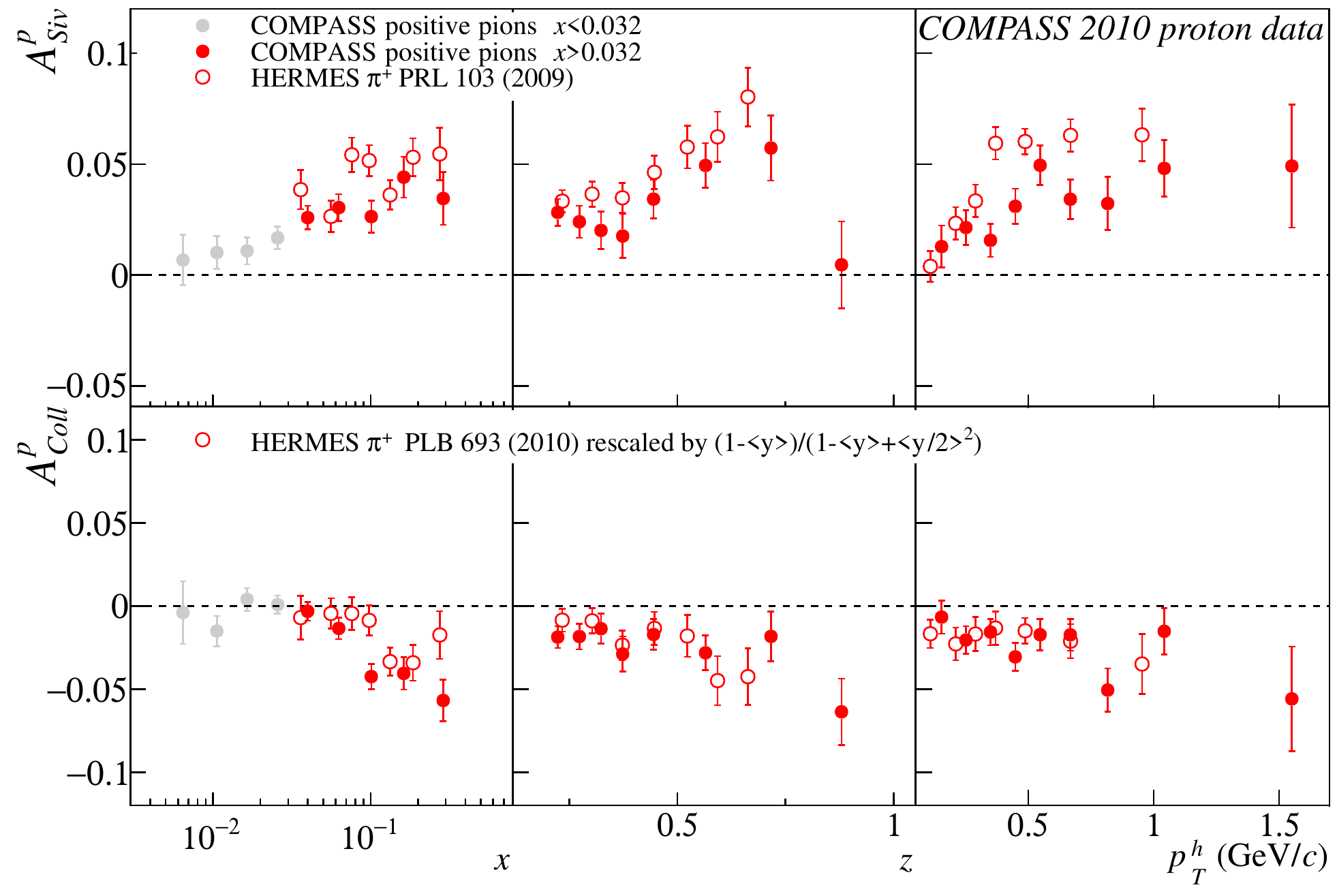}
\caption{Sivers (top) and Collins (bottom) asymmetries at COMPASS and HERMES. \label{fig:ColSivHC}}
\end{figure}
Presently, one of the related challenges in the field of spin-physics is the study of TMD evolution of various PDFs and FFs
and related asymmetries. Comparison of COMPASS and HERMES results for Sivers and Collins asymmetries on proton \cite{Adolph:2014zba,Airapetian:2009ae,Airapetian:2010ds}
emphasized the importance of this domain. In the Figure~\ref{fig:ColSivHC} are demonstrated Sivers (top) and Collins (bottom) asymmetries for positive hadrons
as measured at COMPASS and HERMES experiments. While results for the Collins asymmetries appear to be compatible, Sivers effect at COMPASS at large $x$
is noticeably smaller than the one obtained by HERMES. Here the important detail is that at given $x$ COMPASS $Q^2$ values are
by a factor of $3–-4$ larger than the HERMES ones. Thus, observed
behaviour of Sivers and Collins effects can be used to adjust description of the $Q^2$-dependence of TMDs.
Presently different models predict from small up to quite large QCD-evolution
effects attempting to describe available experimental observations and make predictions for the future
ones \cite{Aybat:2011ta}--\cite{Sun:2013hua}. Additional precise experimental measurements exploring different
$Q^2$ domains for fixed $x$-range are necessary to further constrain the theoretical models.
The work described in this review is a unique and first ever attempt to explore behaviour of TSAs in the multivariate
kinematical environment of the data collected by a single experiment.
For this purpose COMPASS experimental data was split into five different $Q^2$ ranges giving an opportunity to study
asymmetries as a function of $Q^2$ at fixed bins of $x$. Additional variation of $z$ and $p_T$ cuts allows to deeper
explore multi-dimensional behaviour of the TSAs and their TMD constituents.

\section{Multidimensional analysis of TSAs}	
During the "phase-1" (from 2002 to 2010) COMPASS has performed series of SIDIS data-takings using 160 GeV/c
longitudinally polarized muon beam and transversely polarized $^6LiD$ and $NH_3$ targets (See \cite{Adolph:2012sn}--\cite{Parsamyan:2007ju} and references therein).
In 2012 COMPASS entered in "phase-II" and recently performed Drell-Yan measurements with 190 GeV/c $\pi^-$ beam and unpolarized (in 2014) and
transversely polarized $NH_3$-targets (in 2015)
\cite{Parsamyan:2015cfa,Gautheron:2010wva}.

Very soon both sets of COMPASS results from SIDIS and Drell-Yan will become a subject of global fits and phenomenological analyses.
In order to do provide relevant input for these studies, COMPASS SIDIS proton 2010 data has been re-analyzed in a
more differential way extracting the asymmetries in the same four $Q^2$
kinematic regions which were selected for the COMPASS Drell-Yan measurement program \cite{Parsamyan:2015cfa,Gautheron:2010wva}:
$Q^{2}/(GeV/c)^2$ $\in$ $[1;4],[4;6.25],[6.25;16],[16;81]$.
Preliminary results obtained with this selection have been presented in \cite{Parsamyan:2014uda,Parsamyan:2015cfa}
while current review is dedicated to more recent $x$-$z$-$p_T$-$Q^2$ multi-dimensional approach \cite{Parsamyan:2015dfa}.

The analysis was carried out on the same data-sample collected in 2010 with transversely polarized proton target. General event
selection as well as asymmetry extraction and systematic uncertainty evaluation procedures
were identical to those used for recent COMPASS results on Collins, Sivers and other TSAs \cite{Adolph:2012sn}--\cite{Parsamyan:2007ju}.
\begin{figure}[H]
\includegraphics[width=0.5\textwidth]{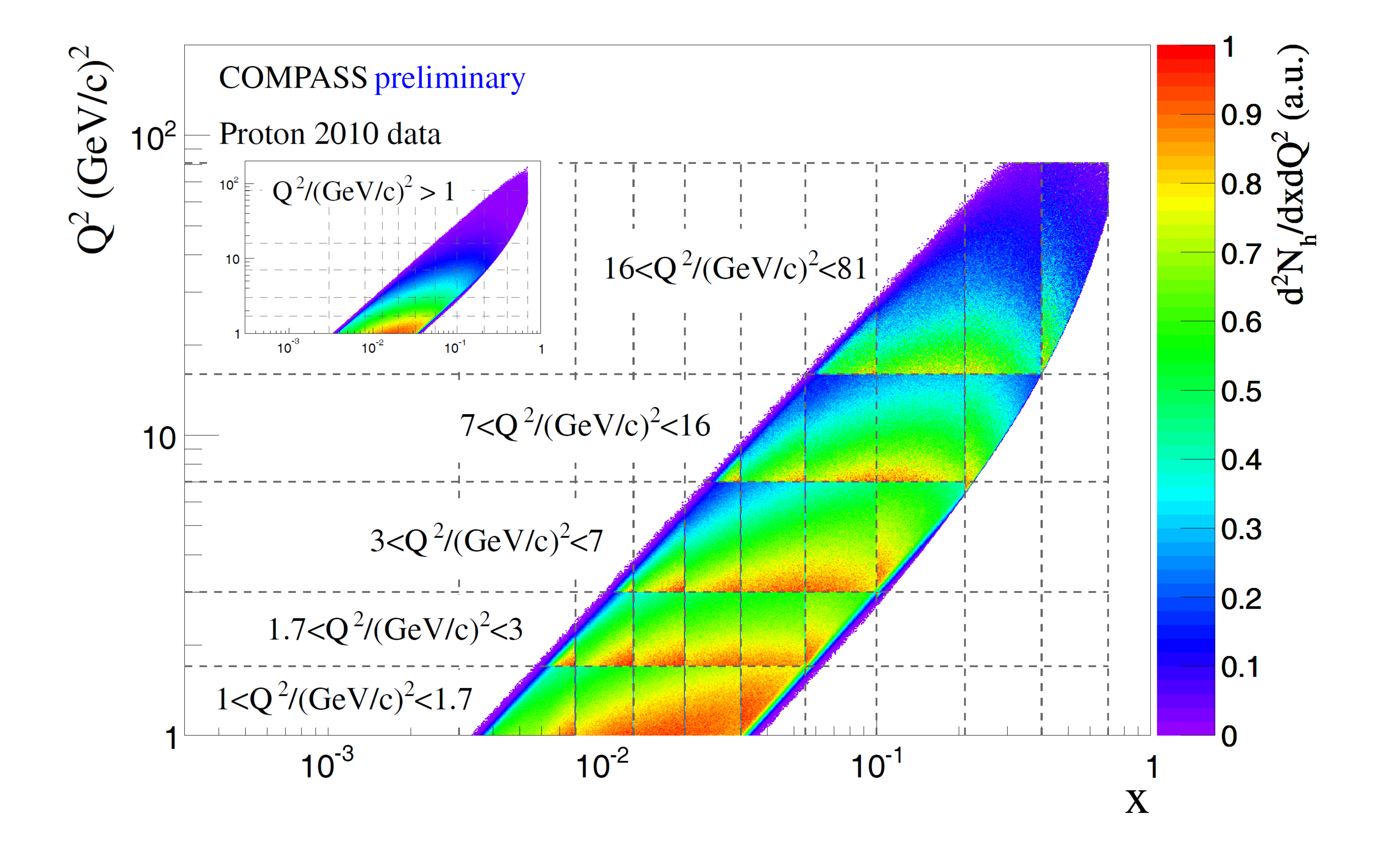}
\includegraphics[width=0.5\textwidth]{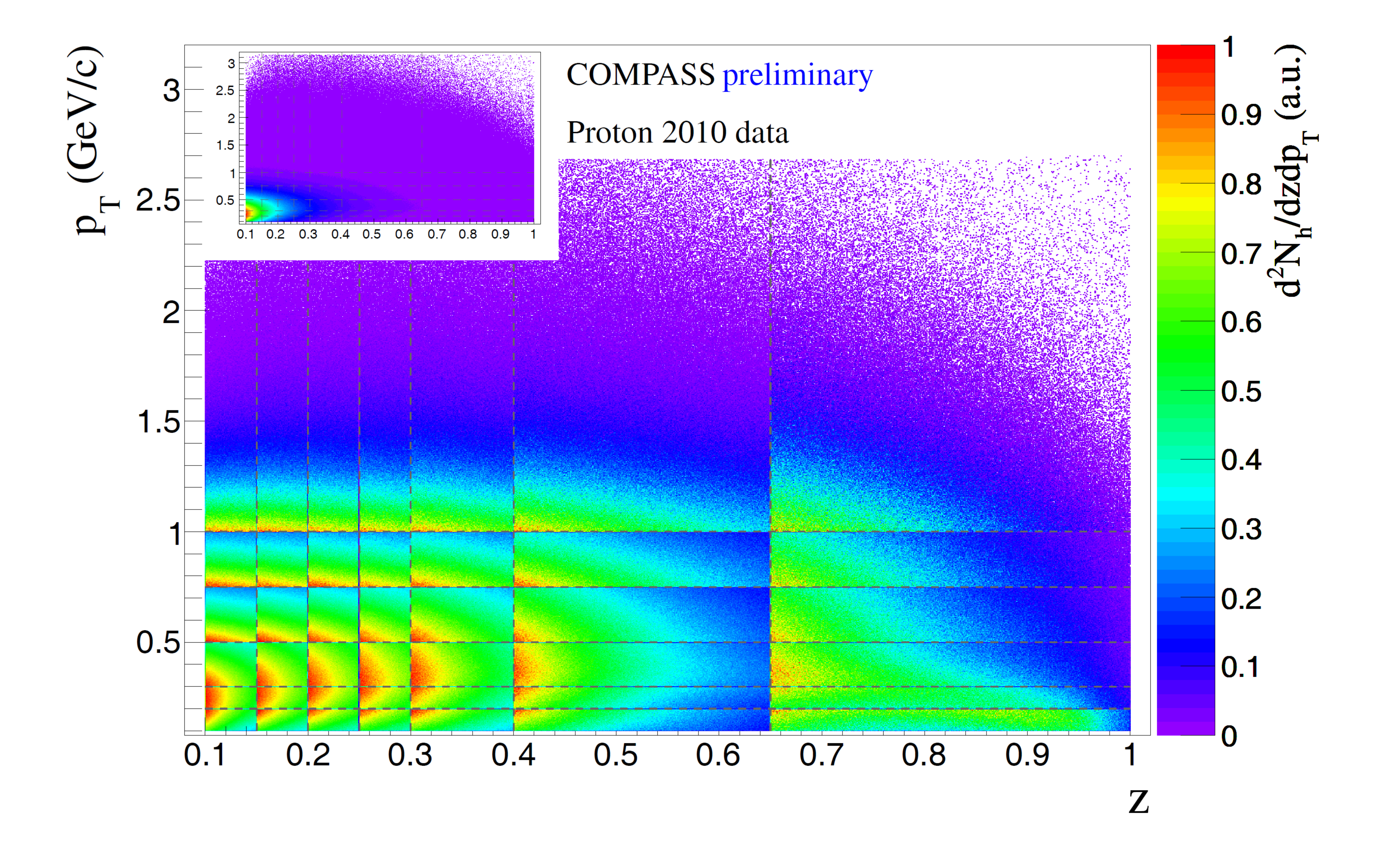}
\caption{COMPASS $x:Q^2$ (left) and $z$:$p_T$ (right) phase space coverage. \label{fig:f1}}
\end{figure}
Amplitudes for all ten azimuthal modulations present in the transverse spin dependent part of the cross-section \ref{eq:SIDIS} have been extracted simultaneously
using extended unbinned maximum likelihood estimator. Obtained "raw" asymmetries have been then corrected for average depolarization factors ($\varepsilon$-depending
factors in equation \ref{eq:SIDIS} standing in front of the amplitudes), dilution factor and target and beam (only
DSAs) polarizations evaluated in the given kinematical bin \cite{Adolph:2012sn}--\cite{Parsamyan:2007ju}.
%
Primary sample is defined by the following standard DIS cuts: $Q^2>1$ $(GeV/c)^2$, $0.003<x<0.7$ and $0.1 <y < 0.9$ and
two more \textit{hadronic} selections: $p_T>0.1$ GeV/c and $z>0.1$.

In order to study possible $Q^2$-dependence the $x$:$Q^2$ phase-space covered by COMPASS experimental
data has been divided into $5\times9$ two-dimensional grid (see left plot in Figure~\ref{fig:f1}).
where five $Q^2$-ranges are the following ones: $Q^{2}/(GeV/c)^2$ $\in$ $[1;1.7],[1.7;3],[3;7],[7;16],[16;81]$.
In addition, each of this samples has been divided into five $z$ and five $p_T$ (GeV/c) sub-ranges defined as follows:\\
$z>0.1$, $z>0.2$, $0.1<z<0.2$, $0.2<z<0.4$ and $0.4<z<1.0$\\
$p_T>0.1$, $0.1<p_T<0.75$, $0.1<p_T<0.3$, $0.3<p_T<0.75$ and $p_T>0.75$.

Using various combinations of aforementioned cuts and sub-ranges, asymmetries have been studied in following "3D" and "4D" configurations:
1) $x$-dependence in $Q^2$-$z$ and $Q^2$-$p_T$ grids.
2) $Q^2$-dependence in $x$-$z$ and $x$-$p_T$ grids. 3) $Q^2$- (or $x$-) dependence in $x$-$p_T$ (or $Q^2$-$p_T$)
grids with different choices of $z$ sub-range.

The second general approach was defined to focus on $z$- and $p_T$-dependences in different $x$-ranges.
 For this study, two-dimensional $z$:$p_T$ phase-space has been divided into $7\times6$ grid
as it is demonstrated in the right plot in Figure~\ref{fig:f1}.
Selecting in addition three different $x$-ranges: $0.003<x<0.7$, $0.003<x<0.032$, $0.032<x<0.7$, asymmetries
have been extracted in "3D: $x$-$z$-$p_T$" grids.
In the next section selected COMPASS preliminary results obtained for multi-dimensional
target transverse spin dependent azimuthal asymmetries are presented.
\section{Results}
\begin{figure}[H]
\center
\includegraphics[width=1\textwidth]{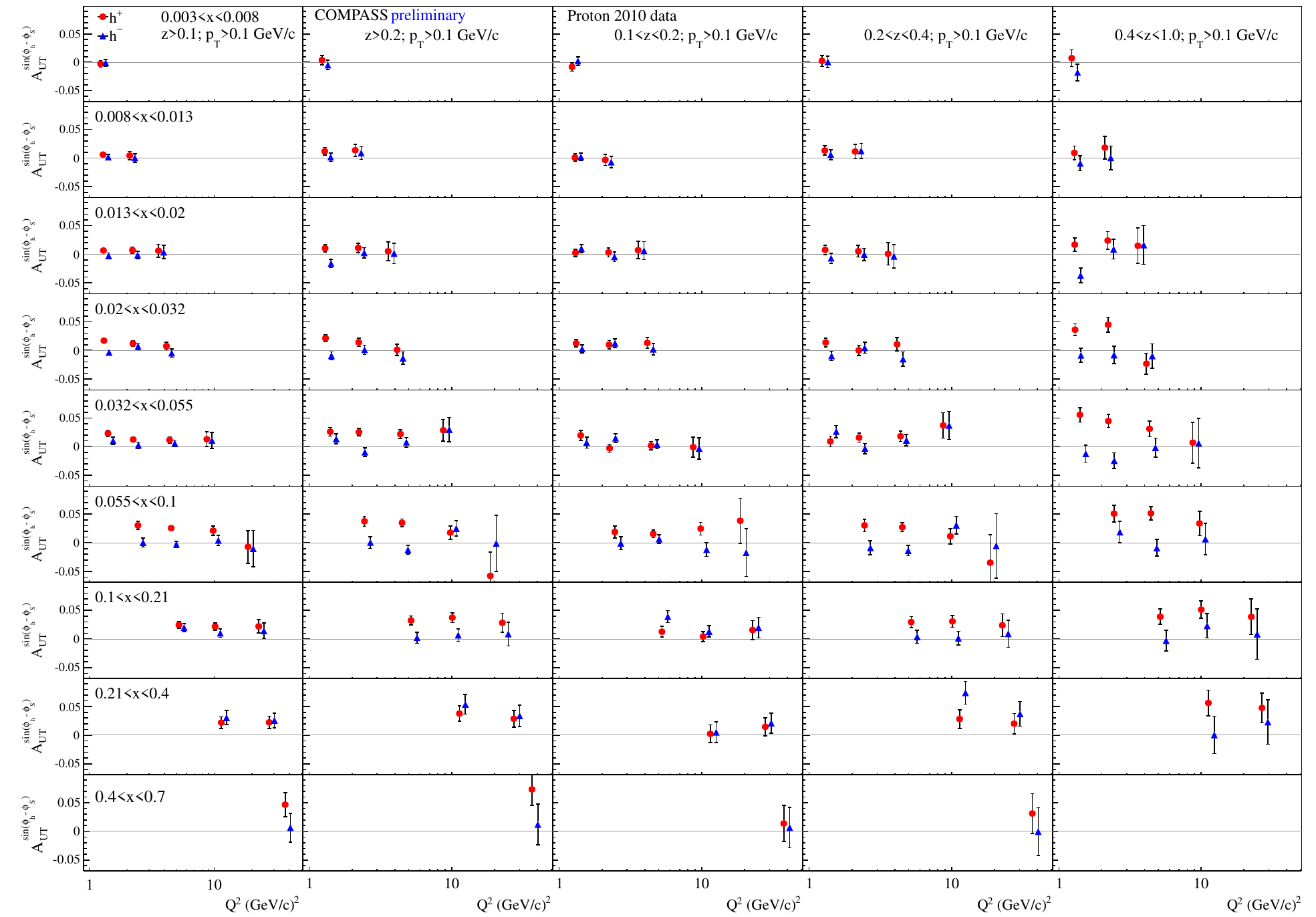}
\includegraphics[width=1\textwidth]{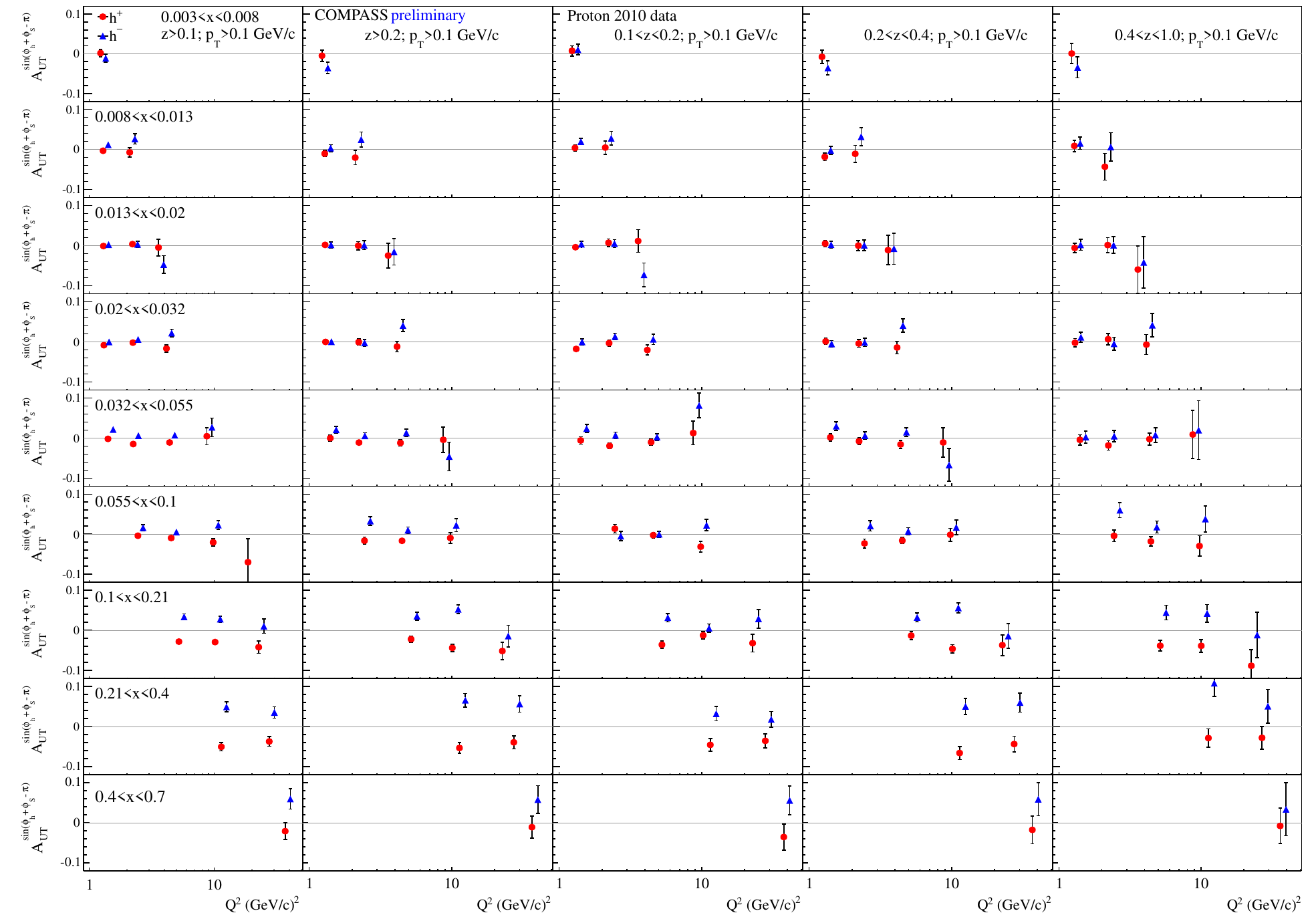}
\caption{Sivers (top) and Collins (bottom) asymmetries in "3D" $x$-$z$-$Q^2$. \label{fig:f2}}
\end{figure}

\footnote{Results discussed in this section have been first presented at the SPIN-2014 conference \cite{Parsamyan:2015dfa},
see also \cite{Parsamyan:2015myq},\cite{Parsamyan:DSPIN15}.}
As an example of "3D" Sivers and Collins effects, results for the extracted $x$-$z$-$Q^2$ configurations are presented in the
Figure~\ref{fig:f2} ('top' and 'bottom' plots, respectively).
As a general observation, for positive hadron production Sivers asymmetry shows sizable signal along whole $x$-range,
while for negative hadrons effect is not clear. Still, there are some indications for a positive Sivers signal at
relatively large $x$ and $Q^2$ and for a negative effect at low $x$.
Clear "mirrored" behaviour for positive and negative hadron amplitudes is being observed in most of the bins for Collins effect.
In general, both Sivers and Collins amplitudes tend to increase in absolute value with $z$ and $p_T$.

Demonstrated in the Figure~\ref{fig:f2} $Q^2$-dependences of Sivers and Collins asymmetries serve as a direct input for TMD-evolution related studies.
In fact, for Sivers effect in several x-bins there are some hints for possible decreasing $Q^2$-dependence for positive hadrons which become more
evident at large $z$. In the meantime, Collins asymmetry does not show any clear indications for $Q^2$-dependence. Thus, both these
observations are in agreement with quoted previously COMPASS-HERMES comparison for Sivers and Collins effects (see \cite{Adolph:2014zba,Airapetian:2009ae,Airapetian:2010ds} and Figure~\ref{fig:ColSivHC}).

\begin{figure}[H]
\center
\includegraphics[width=1\textwidth]{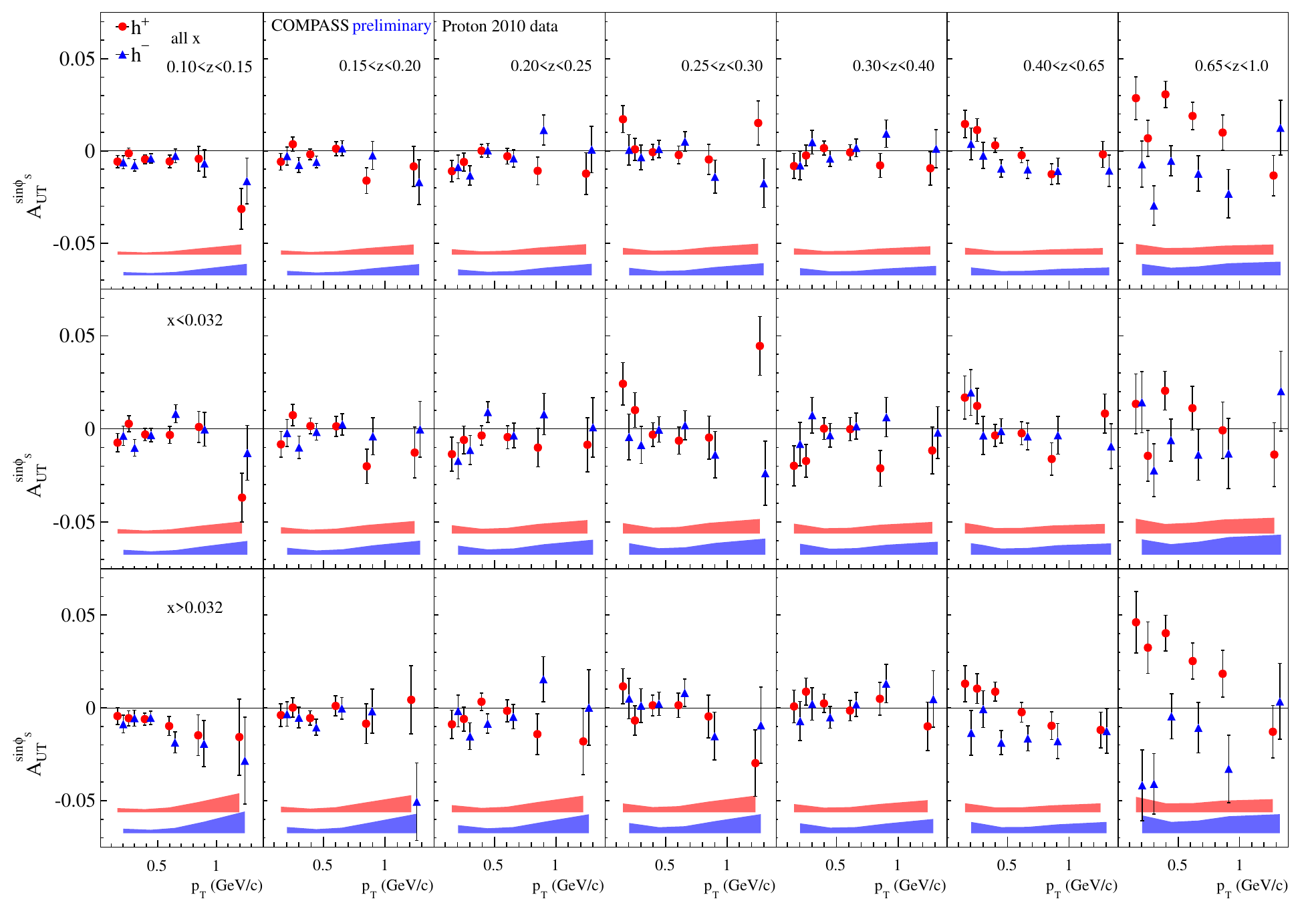}
\includegraphics[width=1\textwidth]{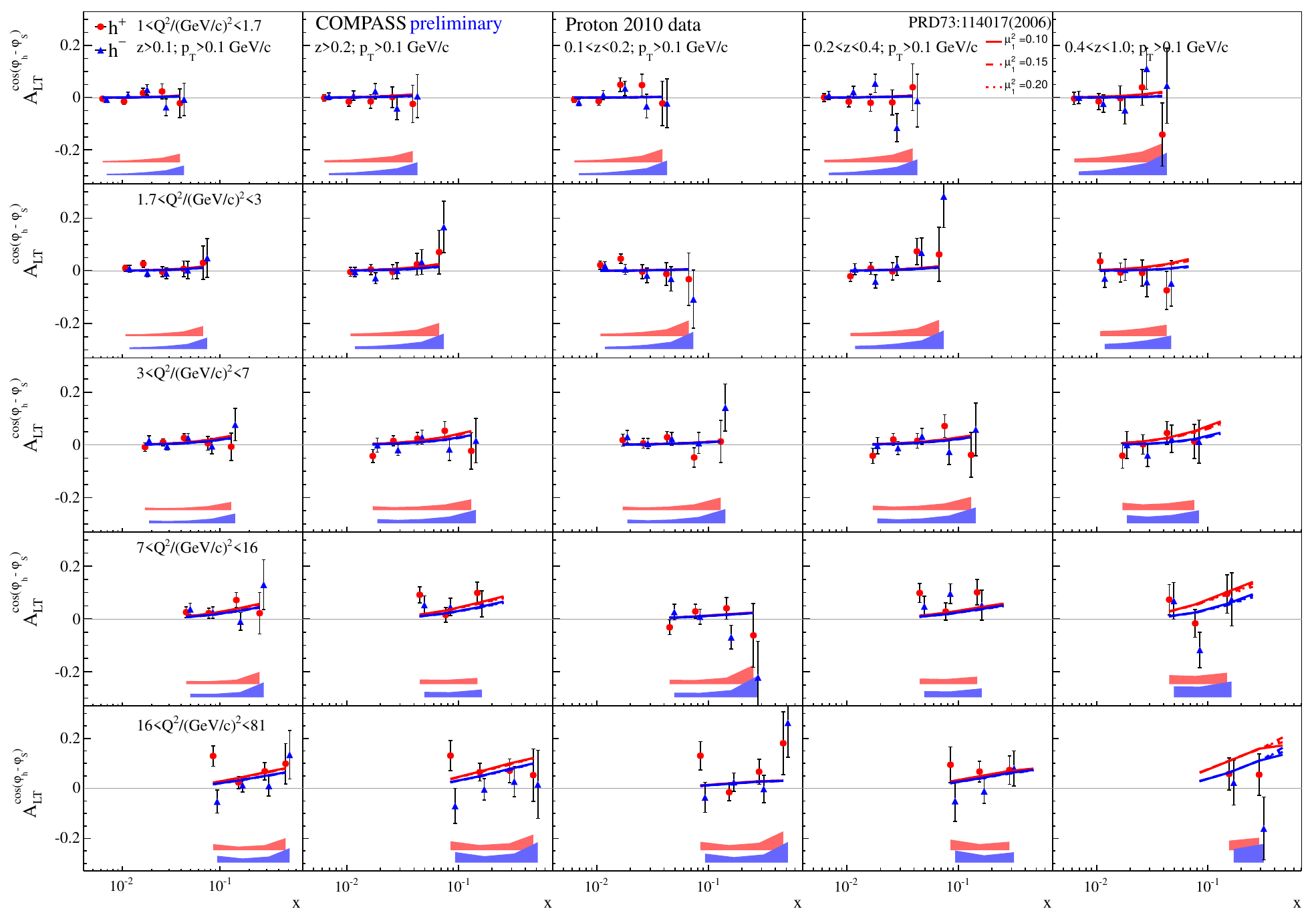}
\caption{Top: $A_{UT}^{\sin (\phi _s )}$ asymmetry in "3D" ($x$-$z$-$p_T$). Bottom: $A_{LT}^{\cos
(\phiH -\phiS )}$ in "3D" ($Q^2$-$z$-$x$) superimposed with theoretical predictions from \cite{Kotzinian:2006dw}.
\label{fig:f3}}
\end{figure}

Another SSA which is found to be non-zero at COMPASS is the $A_{UT}^{\sin (\phi _s )}$ term which is presented in
Figure~\ref{fig:f3} (top) in "3D: $x$-$z$-$p_T$" configuration. Here the most interesting is the large $z$-range were
amplitude is measured to be sizable and non zero both for positive and negative hadrons. It is relevant to remind that within
the "Wandzura-Wilczek approximation" this asymmetry can be associated with Sivers and Collins mechanisms.

\begin{figure}[H]
\center
\includegraphics[width=1\textwidth]{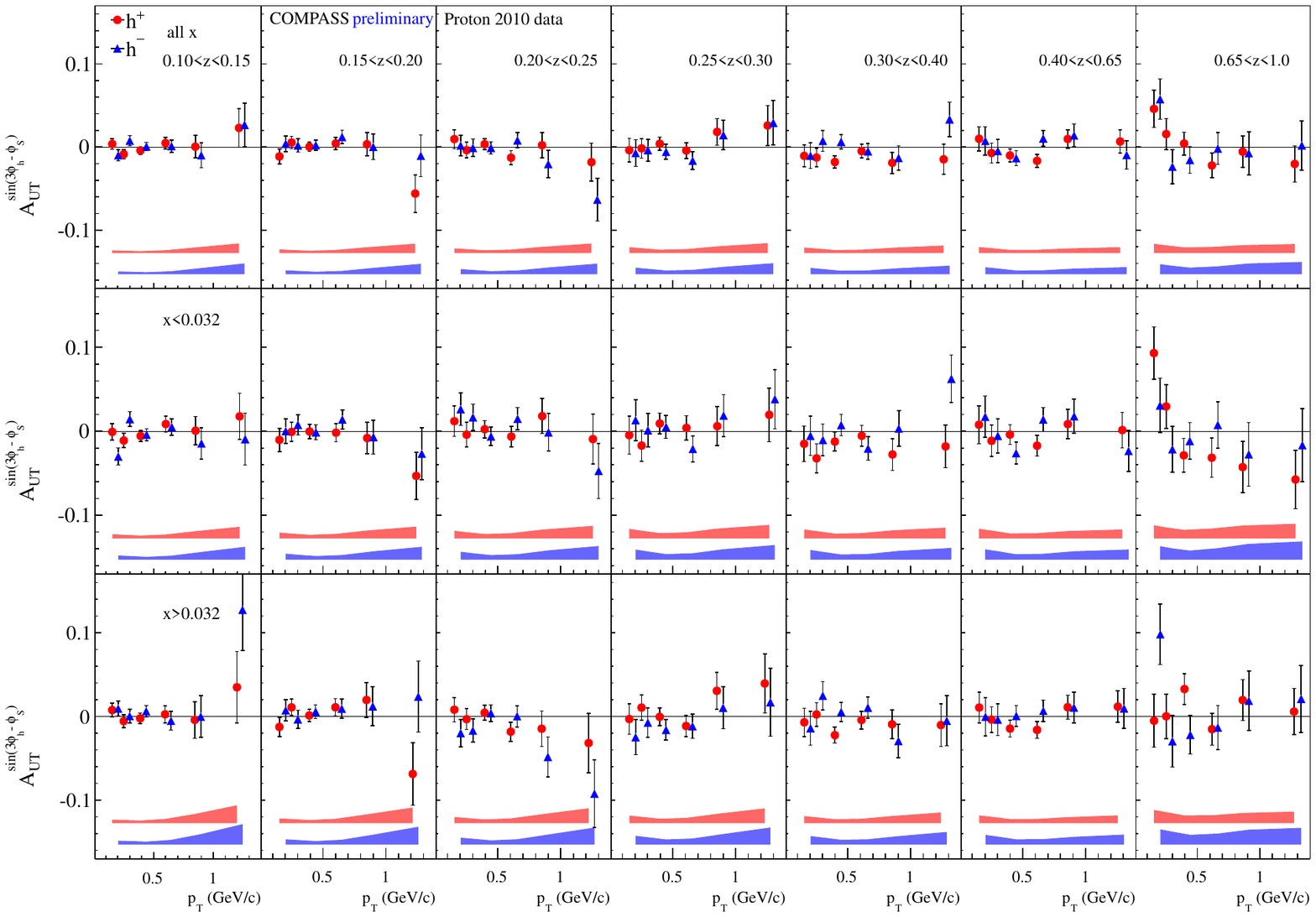}
\caption{$A_{UT}^{\sin (3\phi _s -\phiS)}$ asymmetry in "3D" ($x$-$z$-$p_T$).
\label{fig:f4}}
\end{figure}

The bottom plot in the Figure~\ref{fig:f3} is dedicated to the $A_{LT}^{\cos (\phiH -\phiS )}$ DSA explored in "3D:
$Q^2$-$z$-$x$" grid and superimposed with the theoretical curves from \cite{Kotzinian:2006dw}. This is the only DSA
which appears to be non-zero at COMPASS and the last TSA for which a statistically significant signal has been
detected. Remaining four asymmetries are found to be small or compatible with zero within available statistical
accuracy which is in agreement with available predictions \cite{Mao:2014aoa,Mao:2014fma,Lefky:2014eia}.
As an example the "3D: $x$-$z$-$p_T$" results for $A_{UT}^{\sin (3\phiH -\phiS )}$ asymmetry
are presented in Figure~\ref{fig:f4}.

%
%
%
%
%
%
%
%
%
\section{Conclusions}
COMPASS multidimensional analysis of the whole set of proton TSAs has been performed exploring various multi-differential configurations
in the $x$:$Q^2$:$z$:$p_T$ phase-space. Particular attention was focused on the possible $Q^2$-dependence of
asymmetries, serving a direct input to TMD-evolution related studies.
Several interesting observations have been made studying the results obtained for Sivers,
 Collins, $A_{LT}^{cos(\phi_h-\phi_S)}$ and $A_{UT}^{sin(\phi_S)}$ asymmetries. Other four asymmetries were found to be
compatible with zero within available statistical accuracy.

This is the first ever attempt to extract multi-differential dependencies of all possible transverse spin dependent asymmetries using
experimental data collected by a single experiment. Provided highly differential data set,
combined with past and future relevant data of other collaborations, will give a unique opportunity to access
the whole set of TMD PDFs and test their multi-differential nature.

Also particularly interesting will be the planned comparison of presented SIDIS TSAs with the Drell-Yan asymmetries which soon will be extracted from
first ever polarized Drell-Yan data collected by COMPASS in 2015.
This unique opportunity to explore nucleon spin-structure via two different processes measured with the same experimental setup, will be
the first direct chance to test the universality and key features of TMD PDFs such as, for instance, expected "sign change" of the Sivers function.
%
%
%
%

\end{document}